\documentclass{iau_FM}
\usepackage{graphicx}

\title[Dust/molecules in extra-galactic PNe] 
{Dust and molecules in extra-galactic planetary nebulae}

\author[D. A. Garc\'{\i}a-Hern\'andez]   
{D. A. Garc\'{\i}a-Hern\'andez$^{1,2}$\footnote{D.A.G.H. acknowledges support
provided by the Spanish Ministry of Economy and Competitiveness (MINECO) under
grants AYA$-$2011$-$27754 and AYA$-$2014$-$58082-P.}}

\affiliation{$^1$Instituto de Astrof\'{\i}sica de Canarias, C/ Via L\'actea
s/n, E$-$38205 La Laguna, Spain \\ email: {\tt agarcia@iac.es} \\[\affilskip]
$^2$Departamento de Astrof\'{\i}sica, Universidad de La Laguna (ULL), 
E$-$38206 La Laguna, Spain}

\pubyear{2015}
\jname{Astronomy in Focus, Volume 1} 
\editors{Piero Benvenuti, ed.}
\begin{document}

\maketitle

\begin{abstract}
Extra-galactic planetary nebulae (PNe) permit the study of dust and molecules in
metallicity environments other than the Galaxy. Their known distances lower the
number of free parameters in the observations vs. models comparison, providing
strong constraints on the gas-phase and solid-state astrochemistry models.
Observations of PNe in the Galaxy and other Local Group galaxies such as the
Magellanic Clouds (MC) provide evidence that metallicity affects the production
of dust as well as the formation of complex organic molecules and inorganic
solid-state compounds in their circumstellar envelopes. In particular, the lower
metallicity MC environments seem to be less favorable to dust production and the
frequency of carbonaceous dust features and complex fullerene molecules is
generally higher with decreasing metallicity. Here, I present an observational
review of the dust and molecular content in extra-galactic PNe as compared to
their higher metallicity Galactic counterparts. A special attention is given to
the level of dust processing and the formation of complex organic molecules
(e.g., polycyclic aromatic hydrocarbons, fullerenes, and graphene precursors)
depending on metallicity.

\end{abstract}

\firstsection 
\section{Introduction}

At the end of the asymptotic giant branch (AGB) phase, low- and
intermediate-mass stars experience thermal pulses and strong mass loss. The
strong mass loss efficiently enriches the interstellar medium (ISM) with
specific isotopes and dust grains. The main processes of nucleosynthesis take
place during the thermal pulsing phase on the AGB, while the molecular processes
are more important in the transition phase between AGB stars and planetary
nebulae (PNe). 

During the thermal pulsing phase on the AGB, $^{12}$C and heavy s-process
elements such as Rb, Zr, etc. are produced and dredge-up to the stellar surface.
At solar metallicity, low-mass AGB stars (M$<$1.5 M$_{\odot}$) are O-rich and
they probably do not form PNe. Intermediate-mass AGBs (1.5$<$M$<$4 M$_{\odot}$)
are C-rich and they do not experience hot bottom burning (HBB), while the
high-mass (M$>$4 M$_{\odot}$) AGB stars remain O-rich because of the HBB
activation. This evolutionary scenario has a strong dependence with metallicity
(e.g., Garc\'{\i}a-Hern\'andez et al. 2006, 2007, 2009 and references therein).
For example, the mass limit for HBB activation decreases with decreasing
metallicity. In short, the more massive AGB stars produce different elements
than lower mass AGBs and this should be reflected in the gas and circumstellar
dust chemistry. 

AGB stars are also wonderful molecular factories and more than 60 molecules have
been detected in the circumstellar envelopes around AGB stars (e.g., Herbst \&
van Dishoeck 2009). These molecules are mainly detected via their rotational
transitions from the infrared (IR) to milimeter wavelegnths. They are mainly
gas-phase molecules, including: inorganics such as SiO, SiS, and NH$_{3}$,
organics like C$_{2}$H$_{2}$ and CH$_{4}$, radicals (e.g., HCO$^{+}$), rings
(e.g., C$_{3}$H$_{2}$) or chains (e.g., HC$_{9}$N). Note that gas-phase
reactions can explain most of these molecules but not all of them and
solid-state chemistry has to be considered in the models; e.g., molecules may
form on dust grains.

The composition of the dust around evolved stars depend on the dominant stellar
chemistry. The O-rich AGB stars display amorphous silicates, weak crystalline
silicates like olivine and piroxenes, and refractory oxides such as corundum and
spinel (e.g., Waters et al. 1996), while the C-rich AGBs show silicon carbide
and amorphous carbon, but other complex organic compounds with mixed
aromatic/aliphatic structures (e.g., kerogen) may provide also the strong dust
continuum emission (e.g., Kwok 2004). On the other hand, young and evolved PNe
are mainly characterized by the aromatic infrared bands (usually associated with
polycyclic aromatic hydrocarbons; PAHs) and unidentified infrared emission (UIR)
and crystalline silicates for the case of a C-rich and O-rich chemistry,
respectively. PNe showing both C-rich and O-rich dust features,
what we call mixed-chemistry, are also observed.

PNe in our own Galaxy are characterized by a strong IR dust continuum emission
and the following {\it Spitzer Space Telescope} dust types can be defined
depending on the nature of the solid-state features atop of the continuum
(Stanghellini et al. 2012; Garc\'{\i}a-Hern\'andez \& G\'orny 2014): i)
Featureless (F) PNe with only nebular emission lines in their spectra and little
dust continuum; ii) Carbon chemistry (CC) PNe with C-rich dust features that can
be aromatic and/or aliphatic; iii) Oxygen chemistry (OC) PNe with O-rich dust
features that can be crystalline and/or amorphous; iv) Double-chemistry (DC) PNe
with mixed C-rich and O-rich dust features that can be crystalline and/or
amorphous. Figure 1 (left panel) displays several {\it Spitzer} spectra of
compact Galactic PNe with different dust types. There is an aromatization or
crystallization process from AGB stars to PNe and the aromatic (C-rich) and
crystalline (O-rich) compounds are synthesized during this short transition
phase (Garc\'{\i}a-Hern\'andez 2012 and references therein). Thus, post-AGB
stars or proto-PNe are wonderful laboratories for astrochemistry because they
provide strong constraints on the gas-phase and solid-state chemical models. A
major challenge is to understand the formation pathways of these complex organic
molecules and inorganic solid-state compounds.

\begin{figure}[b]
\begin{center}
 \includegraphics[width=2.4in]{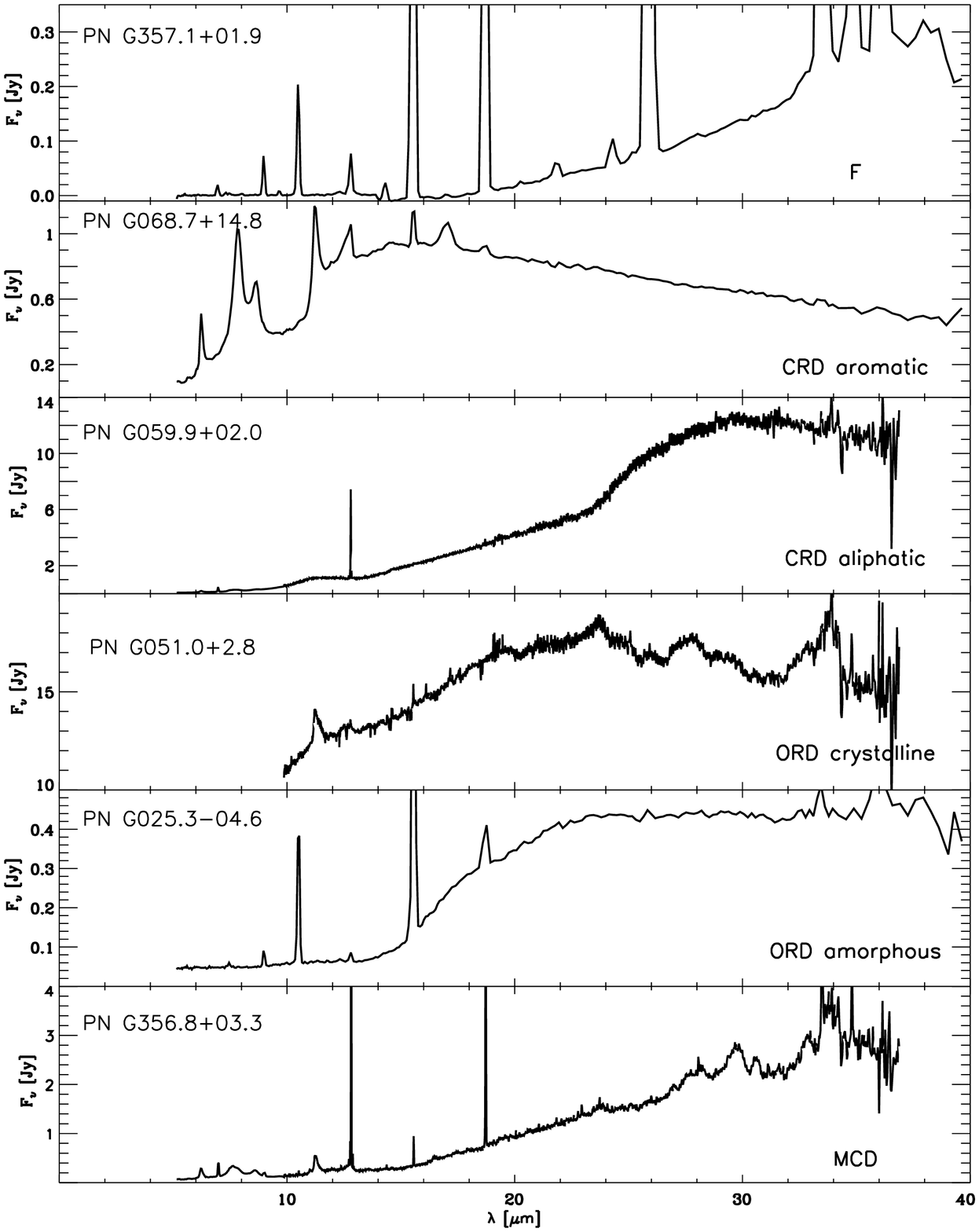}%
 \includegraphics[width=2.3in]{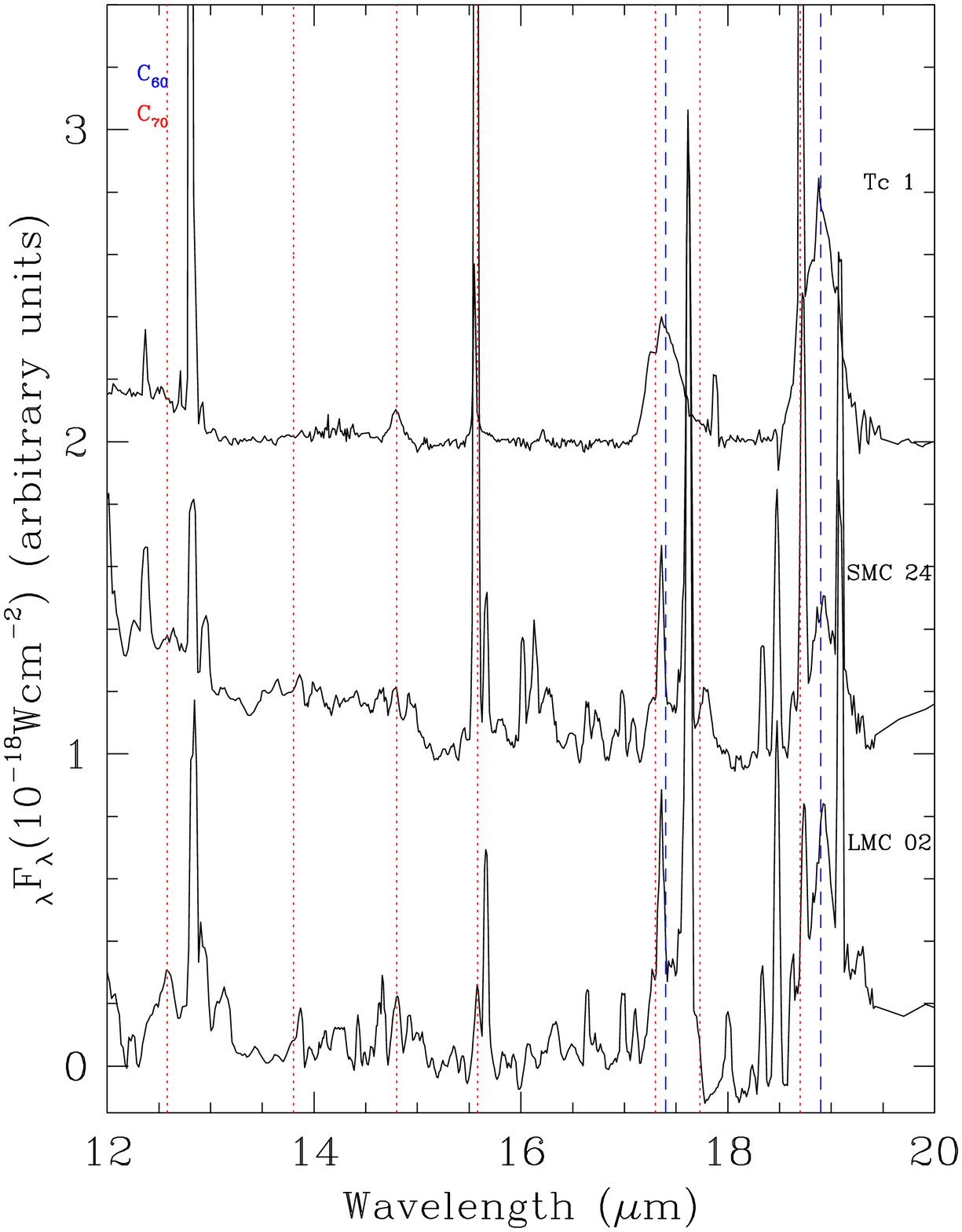} 
 \caption{Left panel: {\it Spitzer} spectral templates of the several dust types
(from top to bottom: featureless (F); carbon-chemistry aromatic or aliphatic
(CC$_{ar}$ or CC$_{al}$), oxygen-chemistry crystalline or amorphous (OC$_{cr}$
or OC$_{am}$), and double-chemistry (DC)), found in compact Galactic PNe
(updated from Stanghellini et al. 2012). Right panel: Residual {\it Spitzer}
spectra ($\sim$10$-$23 $\mu$m) for the C$_{70}$ MC PNe LMC 02 and SMC 24 in
comparison with the Galactic PN Tc 1. The C$_{60}$ (dashed) and C$_{70}$
(dotted) band positions are marked (udapted from Garc\'{\i}a-Hern\'andez et al.
2011).}
   \label{fig1}
\end{center}
\end{figure}

PNe have been detected in sixteen Local Group members. There are dwarf galaxies
nearer to the Milky Way but only a few PNe have been found in them (e.g., Reid
2012). The Magellanic Clouds (MCs) provide single environments at known
distances with higher luminosities than those of the closer galaxies, being an
excellent laboratory for the AGB theories and the molecular and dust formation
models. In addition,  the MCs offer lower metallicity environments than our own
Galaxy. Its relative vicinity provide us with the most complete PNe sample (and
data) with the extra of a low-extinction. In particular, the well known distances
lower the number of free parameters in the comparison between observations and
the models. The AGB nucleosynthesis (and so the dust formation) strongly depend
on metallicity, something that should be reflected in the gas and dust
composition and that we can test by studying PNe in the MCs.

\section{PNe dust types versus metallicity}

The several {\it Spitzer} dust types in PNe vary with the metallicity of the
environment. This is shown in Table 1, where the dust type statistics for
low-metallicitiy MC PNe are presented in comparison with other Galactic
environments such as the Galactic Disk (intermediate metallicity) and Bulge
(high metallicity). The dust features atop the strong dust continua are more
frequent in metal-rich environments like the Galactic Disk and Bulge, as
expected. Within the PNe with solid-state features, carbon chemistry (CC) is
predominant in the MCs, where the crystalline silicates and other O-rich dust
features are very rare (Stanghellini et al. 2007). Also,  double-chemistry (DC)
is not observed in the MCs. The O-rich (OC) and C-rich (CC) objects are equally
predominant in Galactic Disk PNe (Stanghellini et al. 2012;
Garc\'{\i}a-Hern\'andez \& G\'orny 2014), but the Galactic Bulge PNe show a
majority of double-chemistry (DC) dust types (Perea-Calder\'on et al. 2009).

\begin{table}
  \begin{center}
  \caption{Dust type statistics in different metallicity environments.}
  \label{tab1}
 {\scriptsize
  \begin{tabular}{|l|c|c|c|}\hline 
{\bf Dust type} & {\bf Galactic Bulge} & {\bf Galactic Disk} & {\bf Magellanic Clouds}  \\ 
   &  {\bf high z (50)$^{1}$}  & {\bf intermediate z (98)$^{1}$} & {\bf low z (66)$^{1}$} \\ \hline
F          &  14\% (7)    &    19\% (18)    &  41\% (27) \\
CC         &   8\% (4)    &    33\% (32)    &  52\% (34) \\
OC         &  26\% (13)   &    31\% (31)    &   7\& (5)  \\
DC         &  52\% (26)   &    18\% (17)    &   0\% (0)$^{2}$ \\ 
\hline
Ref.$^{3}$ &  S12, GH14   &    S12, GH14    & S07, BS09    \\
\hline
  \end{tabular}
  }
 \end{center}
\vspace{1mm}
 \scriptsize{
 {\it Notes:}\\
  $^1$Total number of PNe in each metallicity environment.\\
  $^2$The LMC and SMC PNe show very similar distributions among the dust types.\\
  $^3$References for the dust types. S12: Stanghellini et al. (2012); GH14: Garc\'{\i}a-Hern\'andez \& G\'orny (2014);
  S07: Stanghellini et al. (2007); BS09: Bernard-Salas et al. (2009).}
\end{table}

{\underline{\it Oxygen-rich chemistry PNe}}. The O-rich chemistry PNe are less
common at the low metallicity of the MCs (Stanghellini et al. 2007) and the
statistics of the {\it Spitzer} dust subtypes (amorphous or crystalline) change
with the metallicity of the environment (e.g., Garc\'{\i}a-Hern\'andez \&
G\'orny 2014). For example, the amorphous silicate features in emission at
$\sim$9.7 $\mu$m are more frequent at the intermediate metallicities of the
Galactic Disk, while the crystalline silicate features (e.g., at 23.5, 27.5 and
33.5 $\mu$m) are very rare in the MCs but they completely dominate at the high
metallicity of the Galactic Bulge. 

{\underline{\it Carbon-rich chemistry PNe}}. In the case of the C-rich PNe, just
the opposite is seen and they are more frequent at the MC low-metallicity (Table
1). The C-rich MC PNe usually show less processed dust grains or small dust
grains (Stanghellini et al. 2007); the aliphatic dust dominates the MC PNe {\it
Spitzer} spectra and the PAHs are not very frequent. Also, the frequency of
C-rich PNe varies with the environment, dominating in the MCs and being absent
in the Galactic Bulge (Table 1). The PNe with aliphatic dust features such as
those at $\sim$9$-$13, 15$-$20, and 25$-$35 $\mu$m (see Section 3) are very
frequent in the MCs, while their Galactic Disk counterparts are usually compact
(and presumably young) PNe of sub-solar metallicity (Garc\'{\i}a-Hern\'andez \&
G\'orny 2014). At solar metallicity, the strong dust continuum emission and the
$\sim$11.5$\mu$m SiC feature seen in the AGB phase evolve to small hydrocarbon
molecules (such as C$_{2}$H$_{2}$) in absorption together with broad 30$\mu$m
emission in the post-AGB stage. The dust features evolve very quickly from
aliphatics to PAHs and when the central star is hot enough to form a PN, the IR
spectrum is dominated by nebular emission lines and we still can see the
30$\mu$m feature (Garc\'{\i}a-Lario \& Perea-Calder\'on 2003). The IR spectral
evolution is a little bit different at the lower metallicity of the MCs and the
transition from aliphatic dust features such as the $\sim$9$-$13 $\mu$m feature
to PAHs seems to be slower (Stanghellini et al. 2007). There exist two main
aromatization models from the AGB phase to the PN stage: i) the dust processing
by the UV photons from the  central star, which changes the structure from
aliphatic to aromatic (e.g., Kwok et al. 2001; Garc\'{\i}a-Lario \&
Perea-Calder\'on 2003); or ii) C$_{2}$H$_{2}$ and its radical derivatives are
the precursors of PAHs (e.g., Cernicharo et al. 2004). 

\section{The unidentified 21, 26, and 30 $\mu$m features}

There is an interesting group of still UIR features located at $\sim$21, 26, and
30 $\mu$m, which are only observed in C-rich stars evolving from the AGB phase
to the PN stage. 

The 21 $\mu$m feature is usually observed in proto-PNe, indicating that its
carrier is of solid-state origin with a fragile character (see also Section 4).
On the other hand, the 30 $\mu$m feature (sometimes with substructure at
$\sim$26 $\mu$m) is seen all the way from AGB stars to PNe, suggesting that its
carrier is very abundant in the circumstellar envelope. Many carriers for this
specific UIR features have been proposed to date. For example, nanodiamonds,
hydrogenated fullerenes (fullerAnes), and hydrogenated amorphous carbon grains
(HACs) have been linked to the 21 $\mu$m feature, while the 30 $\mu$m feature
could be explained by magnesium sulfide (MgS), aliphatic chains, or HACs
(Garc\'{\i}a-Hern\'andez 2012 and references therein). It is very interesting
that the identification of these UIR features as HAC like-dust (some kind of
solid with a mixed aromatic/aliphatic structure) may explain the detection of
complex molecules such as fullerenes and planar C$_{24}$ (graphene precursors)
in some C-rich PNe (see below).

\section{Complex organic molecules versus metallicity}

Fullerenes such as C$_{60}$ and C$_{70}$ are very stable molecules. These
complex organic molecules are very important for interstellar/circumstellar
chemistry because they may explain many astrophysical phenomena like the diffuse
interstellar bands (DIBs) and the ultraviolet bump. Fullerenes were discovered
in the laboratory by Kroto et al. (1985) and they are also found on Earth and
meteorites. C$_{60}$$^{+}$ was also tentatively detected in the ISM (Foing \&
Ehrenfreund 1994), which has been recently confirmed by laboratory spectroscopy
(Campbell et al. 2015). At laboratory, fullerenes are efficiently produced under
H-poor conditions and in 2010 Cami et al. reported the first IR detection of
neutral C$_{60}$ and C$_{70}$ in the young PN Tc 1 (which shows a lack of PAH
features in its {\it Spitzer} spectrum) as due to the H-poor conditions in the
circumstellar envelope. 

{\underline{\it Fullerenes in Galactic H-rich PNe}}. We now know that fullerenes
can be detected together with PAHs and that they are efficiently formed in
H-rich circumstellar envelopes only (Garc\'{\i}a-Hern\'andez et al. 2010;
Garc\'{\i}a-Hern\'andez, Rao \& Lambert 2011). Fullerenes are detected in
Galactic PNe with normal H abundances and this result is confirmed by the
independent detection of C$_{60}$ in only those R Coronae Borealis stars with
some H. Remarkably, fullerene PNe display broad HAC-like dust features at
$\sim$9$-$13 and 25$-$35 $\mu$m, suggesting that both fullerenes and PAHs
probably evolve from the photochemical processing of HACs, as indicated by some
laboratory experiments (Scott et al. 1997). Indeed, fullerenes have been
detected also in other Galactic objetcs such as a proto-PN, reflection nebulae,
and young stellar objects, and none of these space environments is H-poor (see
e.g., Garc\'{\i}a-Hern\'andez 2012 and references therein).

{\underline{\it Extra-galactic fullerenes}}. Fullerenes have been detected in
PNe of the MCs and the first extra-galactic detection of C$_{70}$ was reported
(Garc\'{\i}a-Hern\'andez et al. 2011). Figure 1 (right panel) shows the C$_{60}$
and C$_{70}$ emission features detected in MC PNe. The combination of the {\it
Spitzer} spectra with laboratory data permitted the determination of accurate
abundances of both C$_{60}$ (0.07\%) and C$_{70}$ (0.03\%). The great variety of
molecular species (HACs, PAHs clusters, fullerenes, etc.) observed in MC PNe
seems to support the HACs scenario proposed by Garc\'{\i}a-Hern\'andez et al.
(2010), where fullerenes may evolve from the UV-induced HACs decomposition (see
also Micelotta et al. 2012).

{\underline{\it Fullerenes detection vs. metallicity}}. Interestingly, the
detection rate of fullerenes in C-rich PNe increases with decreasing
metallicity; $\sim$5\% of fullerene PNe are found in our Milky Way, while the
fullerenes detection rate is $\sim$20\% and $\sim$44\% in the LMC and SMC,
respectively (Garc\'{\i}a-Hern\'andez et al. 2012). This suggests a more limited
dust processing (or the general presence of small dust grains) at low
metallicity. Indeed, Otsuka et al. (2014) have shown that all Galactic fullerene
PNe are sub-solar metallicity low-mass PNe, which demonstrate that low metallcity
environments favours fullerene production and detection. 

In addition, the 21 $\mu$m feature is more common in the MCs than in the Galaxy
(Volk et al. 2011) and its carrier may be related with the formation of
fullerenes. Volk et al. (2011) reported an anti-correlation between the 30
$\mu$m and other UIR features for the MC 21 $\mu$m sources. Such an
anti-correlation could result from radiation-induced decomposition of HAC grains
into PAHs and fullerenes. Note that in the HACs scenario the 21 $\mu$m feature
is also related with the formation of fullerenes and its carrier may be a
fragile intermediate product from the decomposition of HAC or a similar
material. However, a more recent study on a smaller sample of Galactic proto-PNe
claims that for the Galactic 21 $\mu$m sources the 30 $\mu$m and the UIR
features do not seem to be anti-correlated (Mishra et al. 2015). It is to be
noted here that spectroscopically the MC 21 $\mu$m sources sources differ from
the Galactic ones, with the MC 21 $\mu$m sources displaying more typical UIR
features (i.e., more PAH-like).

{\underline{\it Graphene precursors in PNe}}. Unusual emission features at
$\sim$6.6, 9.8, and 20 $\mu$m have been also detected in MC PNe
(Garc\'{\i}a-Hern\'andez et al. 2011). These features are coincident with the
theoretical transitions of the planar C$_{24}$ molecule (just a piece of
graphene or a graphene precursor). However, a confirmation has to wait for
laboratory spectroscopy, which is extremely difficult because of the high
reactivity of C$_{24}$. The most interesting point here is that the possible
detection of graphene precursors (C$_{24}$) opens the possibility of detecting
other forms of C in space. For example, fullerenes and PAHs may
coexist in fullerene PNe and C$_{60}$ can react with a small PAH like anthracene
via Diels-Alder reactions to form fullerene/anthracene mono-adducts or
bis-adducts. Such fullerene/PAHs adducts display the same mid-IR features of
isolated C$_{60}$ molecules and we could not distinguish them via mid-IR
spectroscopy only (Garc\'{\i}a-Hern\'andez, Cataldo \& Manchado 2013)..

\section{Concluding remarks}

In summary, the different dust properties (and evolution) at different
metallicities (MCs, Galactic Disk and Bulge) are consistent with the AGB
nucleosynthesis theoretical predictions (third dredge-up and HBB). The
coexistence of PAHs, fullerenes, HACs, and possible planar C$_{24}$ (graphene
precursosr) suggests a top-down scenario for the fullerenes formation. The dust
processing is more limited at the low metallicity of the MCs, which otherwise
provide the most favorable conditions for fullerene formation and detection. A
complex family of fullerene-based molecules (e.g., fullerene/PAH adducts,
metallofullerenes, etc.) is likely to be present in space but more laboratory
efforts are needed.

\end{document}